# Inelastic Neutron Scattering Studies of the Intermediate Valence Compound CePd$_3$

V. R. Fanelli[1,2,*], J. M. Lawrence[1], C. H. Wang[1,2], A. D. Christianson[3],
E. D. Bauer[2], K. J. McClellan[2], E. A. Goremychkin[4,5] and R. Osborn[4]

[1]*University of California Irvine, Irvine, CA 92697, USA*
[2]*Los Alamos National Laboratory, Los Alamos, NM 87545, USA*
[3]*Oak Ridge National Laboratory, Oak Ridge, TN 37831, USA*
[4]*Argonne National Laboratory, Argonne, IL 60439, USA*
[5]*ISIS Facility, Appleton Rutherford Laboratory, Chilton, Didcot OX11 0QX, United Kingdom*

Inelastic neutron scattering measurements on a CePd$_3$ single crystal show a magnetic response at 300 K that is independent of momentum transfer with a Lorentzian quasielastic energy spectrum with a half width $\Gamma$ = 23 meV. This is in agreement with the Anderson impurity model (AIM), that predicts local moment relaxational behavior in this temperature regime. The 7 K magnetic response has an inelastic Lorentzian spectrum, with characteristic energy $E_0$ = 53 meV and $\Gamma$ = 32 meV at the ($h$, 1/2, 0) zone boundary. Such an inelastic spectrum is expected for the AIM at low temperature. Unlike the $Q$-independence of the impurity model, a variation of intensity with momentum transfer, including intensity maxima at the zone boundary, is observed in the data. However, this variation is only of order 20 percent, which is *much* smaller than that predicted by the Anderson lattice model (ALM). The large shifts in spectral weight expected for the ALM as $Q$ varies from zone boundary to zone center are not observed in the experimental spectra.

In rare earth intermediate valence (IV) compounds, 4$f$ electrons hybridize with conduction electrons in the presence of strong Coulomb correlations [1]. This represents a classic correlated electron problem. It is a general belief that the Anderson lattice model (ALM) should describe the behavior of IV compounds. However, despite the fact that in these materials the 4$f$ electrons sit on a periodic lattice, the Anderson *impurity* model (AIM) does a surprisingly good job in describing the temperature dependence of the susceptibility $\chi$, the linear coefficient $\gamma$ of specific heat, and the $f$-occupation number $n_f$[2]. Many studies of the dissipative (imaginary) part of the dynamic susceptibility $\chi$" of rare earth IV compounds have been performed using neutron scattering on polycrystalline samples [2,3]. These studies show a crossover between a high temperature regime with a quasielastic energy spectrum for $\chi$" (with width on the order of the Kondo temperature $T_K$) and a low temperature regime with an inelastic Lorentzian energy spectrum centered around $T_K$. This behavior is also consistent with the predictions of the AIM.

While the AIM, being an impurity theory, yields no dependence of the dynamic spin susceptibility on the momentum transfer $Q$, calculations using the ALM predict a strong $Q$-dependence. At low temperatures, various approximate treatments [4] of the ALM predict the onset of hybridized bands (Fig. 1 (inset))

$$\omega_{\vec{k}}^{\pm} = \frac{1}{2}\left(\varepsilon_{\vec{k}} + E_f \pm \sqrt{(\varepsilon_{\vec{k}} - E_f)^2 + 4|V|^2}\right). \quad (1)$$

In this renormalized ground state, the $f$ level energy $E_f$ and the matrix element $V$ for mixing between the $f$ and conduction electrons have been reduced to small effective values by the Coulomb correlations. The hybridization gap is directly observed in the optical conductivity $\sigma(\omega)$ of Kondo insulators such as YbB$_{12}$ [5]. Evidence for the existence of the gap in IV metals can be found in the optical conductivity of CePd$_3$ [6] and Yb compounds [7] in a form of a deep minimum in $\sigma(\omega)$ separating a narrow Drude resonance from a mid-infrared (0.1 - 0.2 eV) peak due to excitations across the gap.

The $Q$-dependence of the dynamic susceptibility $\chi$" arises from the $Q$-dependence of the joint density of initial and final states for the particle-hole excitations. In Fig. 1 we show the resulting spectra for the interband transitions, as determined by Aligia and Alascio [8]; calculations by other authors [9-11] give essentially identical results. An inelastic peak in $\chi$" is obtained with a maximum of intensity when the energy transfer equals the threshold for indirect transitions between the regions of large density of states at the zone center and zone boundary of the upper and lower bands respectively. The momentum transfer for this indirect threshold scattering is at the zone boundary. As $Q$ decreases towards zone center the peak decreases in intensity and moves towards higher energy. At zone center, the peak occurs in the mid-infrared, as for the $Q$ = 0 transitions of the optical conductivity, and the scattering is vanishingly small on the energy scale shown in Fig. 1. (Intraband transitions (not shown) have the opposite $Q$-dependence, with peaks moving to larger energies as $Q$ goes from zone center to zone boundary ($E_{peak} \propto Q$).) Scattering that is large at the zone boundary and that decreases in amplitude and moves to higher energy as $Q$ is reduced towards zone center has been observed in Kondo insulators such as TmSe [12] and YbB$_{12}$ [13], where it indeed has been interpreted as interband scattering across a hybridization gap. The point in question is whether similar shift in spectral weight with $Q$ is found in the IV metals.

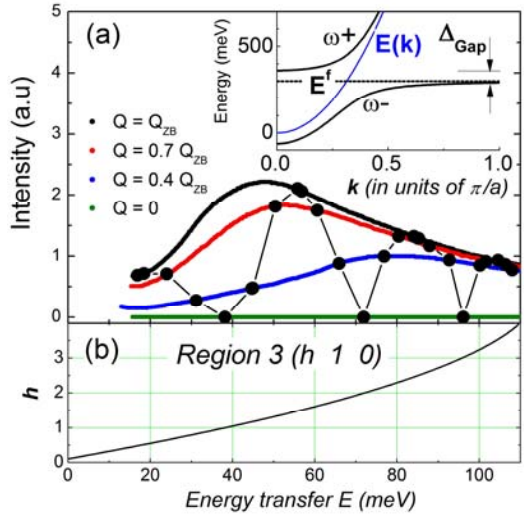

Fig.1: (a) Spectra calculated for the Anderson lattice [8] for interband excitations for momentum transfer in the range between 0 and $Q=Q_{ZB}$. Inset: Showing the hybridized bands $\omega^+$ and $\omega^-$ (solid lines) that arise from a dispersionless $E_f = 300$ meV band (dashed line) that crosses a conduction band (blue line) $E(k) \propto k^2$. (b) The variation of the $h$ component of momentum transfer with energy transfer for the MAPS measurement in region 3. Symbols are drawn on the spectra of (a) at the energies where $h(E)$ takes on the (reduced) value of $Q$ appropriate to each spectrum; the thin line is a smooth interpolation between these symbols.

To look experimentally for such a $Q$-dependence, single crystals are required. In this letter, we present inelastic neutron scattering (INS) measurements on the IV compound CePd$_3$ and LaPd$_3$ under identical conditions. A CePd$_3$ crystal of diameter 0.5 cm and length 5 cm, with a mass of 17.72 g, and a LaPd$_3$ crystal of diameter 0.6 cm and length 3 cm, with a mass of 10.55 g, were prepared by the Czochralski method. The crystals were aligned using the HB1a and HB3 spectrometers at the High Flux Isotope Reactor at the Oak Ridge National Laboratory. Both neutron and X-ray diffraction confirmed that the CePd$_3$ and LaPd$_3$ samples were single phase, with a mosaicity of 2.8 degrees and 3.5 degrees, respectively. INS measurements have been performed at the pulsed spallation neutron source ISIS of the Rutherford Appleton Laboratory on the time-on-flight (TOF) chopper spectrometer MAPS with initial energy of the neutrons fixed at 120 meV and at temperatures of 7 and 300 K, with instrumental energy resolution of 9.5 meV. The sample was oriented with its [100] direction parallel to the incident beam.

To separate the magnetic from the non-magnetic component of the scattering, we assumed that the phonon and multiple scattering contributions in CePd$_3$ are given by the energy spectra of the nonmagnetic counterpart LaPd$_3$ after a proper scaling [14]. In addition to scaling by the ratio of the sample masses, the LaPd$_3$ scattering spectrum per La ion was scaled by a factor of 0.7, a value coming from the ratio of the total scattering cross sections for both compounds. Figure 2(a) shows how the scaled spectrum from LaPd$_3$ matches the one of CePd$_3$ in the energy transfer range of 10 to 20 meV where the single-phonon scattering contribution is dominant. The data shown correspond to a region in reciprocal space that will be defined below. The difference between the CePd$_3$ and the scaled LaPd$_3$ scattering should therefore account for the magnetic scattering, especially above 25 meV, that is, for energy transfers greater than the energy range of single-phonon scattering events. From Fig. 2(a), the magnetic scattering is maximum between 50 and 70 meV at 7 K.

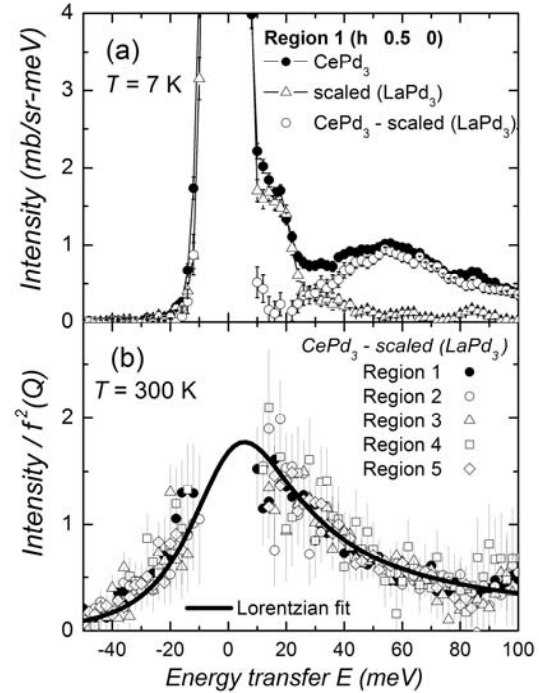

Fig.2: Scattered intensity measured on the MAPS spectrometer with incident neutrons of 120 meV. (a) Scattering at 7 K from the CePd$_3$ sample (black circles), scaled scattering from LaPd$_3$ (open triangles) and the difference (open circles), assumed to represent the magnetic contribution to the scattering. These data correspond to Region 1 in the reciprocal space defined in the text. (b) Magnetic contribution to the scattering normalized by the magnetic form factor at five regions in $Q$-space at 300 K, and the correspondent quasielastic Lorentzian fit (black line).

The distribution of the scattered intensity over momentum transfer space is shown in Fig. 3, where the scattered intensity is integrated over the energy transfer range between 50 and 70 meV corresponding to the maximum of the magnetic response of the system observed at low temperatures (Fig. 2(a)). The scattering is relatively uniform with $Q$ at room temperature (Fig. 3(b)), whereas it shows a variation in intensity around 20 % at low temperatures (Fig. 3(a)), being maximum at zone boundaries ($h$, 0.5, 0).

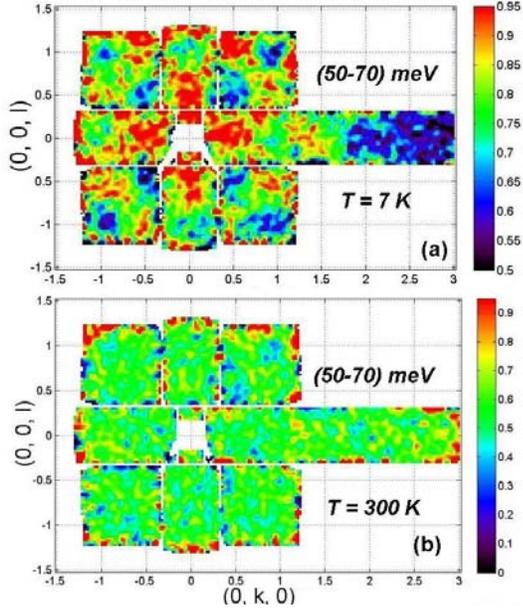

Fig.3: Scattering intensity over the ($k$, $l$) plane integrated over an energy transfer range of 50 to 70 meV, for neutron scattering data taken on CePd$_3$ at 7 K (a), and at 300 K (b), using MAPS spectrometer with incident neutron energy $E_i$ of 120 meV. The color scale gives the intensity in mb/sr-meV units.

We further demonstrate the $Q$-independence of the magnetic scattering at room temperature by comparing five different $Q$-regions in the plane $(Q_K, Q_L) = (2\pi/a_0)(k, l)$. Regions 1, 2, 3, 4, and 5 are centered at $(k, l) =$ (0.5, 0), (0.5, 0.5), (1, 0), (1.5, 0), and (0.5, 1) respectively and have a width of ± 0.15 for each component $k$ and $l$. Fig. 2(b) shows the magnetic part of the intensity normalized by the form factor for the five regions in $Q$-space. All these spectra can be fit simultaneously by a single quasielastic Lorentzian power spectrum whose intensity and halfwidth ($\Gamma$ = 23.3 meV, or equivalently, 280 K) do not depend on the momentum transfer $Q$. In other words, at room temperature the magnetic response is $Q$-independent, characteristic of (independent) spatially localized magnetic moments, and shows a purely relaxational spin dynamics as expected for the uncorrelated local magnetic moment regime at high temperatures. This local moment limit is already achieved at 300 K, roughly half of the Kondo temperature.

The low temperature magnetic contributions to the spectra for the regions 1-3 of reciprocal space are shown in Fig. 4(a, b, and c). In region 1 (3), the magnetic contribution $S_{magn}(\vec{Q}, E)$ at 7 K can be represented by an inelastic Lorentzian power spectrum with characteristic energy $E_0$ =52.6 (37) meV and $\Gamma$=32 (44) meV.

In TOF measurements with a fixed sample orientation, momentum transfer $Q$ and energy transfer $E$ are coupled, and only three of the four variables $E, h, k, l$ are independent [16]. The variation of the $h$-component of $Q$ with energy transfer is also plotted in each panel of Fig. 4 for each of the regions in $Q$-space. While regions 1 and 3 show reasonable agreement with an inelastic Lorentzian lineshape, in region 2 (and possibly also in region 3) there is an oscillation of the magnetic intensity mounted on top of the Lorentzian curve, being of the same nature as the one seen in the intensity color plot of Fig. 3(a). In other words, the variation of $h$ with energy transfer $E$ in a spectrum at fixed ($k, l$) can lead to an intensity oscillation with $E$ similar to the oscillation in intensity observed when $k$ or $l$ traverses reciprocal-space at a fixed energy transfer $\Delta E$. The oscillation observed in region 2 occurs along the ($h$, 1/2, 1/2) zone edge, with minima when $h = 1/2$ in reduced units, i.e. at the zone corners.

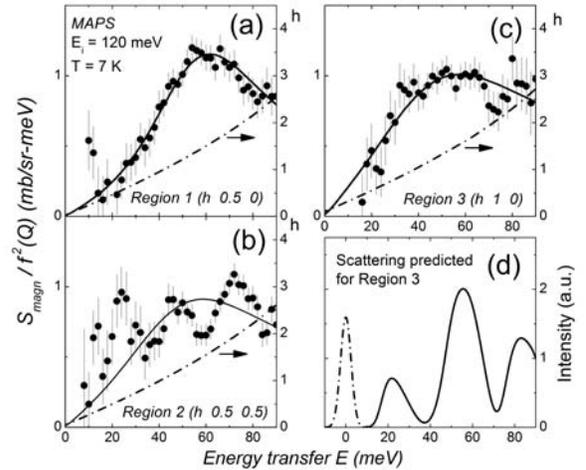

Fig.4: (a,b,c) Magnetic contribution (CePd$_3$ - 0.7 LaPd$_3$) to the intensity spectra normalized by the magnetic form factor (black circles) at three regions in the plane ($k, l$), from measurements on MAPS. The black lines are fits to an inelastic Lorentzian power spectrum, with parameters given in the text. The $h$ component of $Q$ is shown at each region (dashed-dot line). (d) Magnetic contribution to the scattering (solid line) expected for region 3 based on the calculations shown in Fig. 1, after convolution with the instrumental resolution for the MAPS measurements (dashed-dot line).

Based on the spectra of Fig. 1, deep minima are expected in the spectrum for the trajectory ($h(E)$, 1, 0) of region 3 whenever $h(E)= 0$ in reduced units. To show this, we plot symbols on the spectra of Fig. 1(a) wherever $h(E)$ for region 3 takes on the reduced value of $Q$ associated with each individual spectrum. The thin line of Fig. 1(a) interpolates between these points. In Fig. 4(d), we plot this spectrum after convolution with the instrumental resolution of MAPS. In the experimental spectrum for region 3 (Fig. 4(c)), no such deep oscillations are observed; indeed the spectrum for this trajectory differs by less than 20 % from the spectrum of region 1. Hence, while Fig. 3 demonstrates that the $Q$-dependence of the low temperature spin dynamics resembles the threshold interband scattering predicted for the ALM in the sense that intensity maxima occur for zone boundary $Q$, nevertheless closer comparison of Fig. 4 and Fig. 1 shows that the very large shifts in spectral weight with $Q$ expected for the ALM [8,9] are not observed.

If we ignore the 20 percent variation with $Q$, our results for the magnetic response in CePd$_3$ are in good qualitative agreement with calculations by Cox et al [19] for the temperature dependence of the dynamic susceptibility of the Anderson impurity model for intermediate valent CePd$_3$. The AIM predicts an inelastic spectrum $\chi'' \propto \Gamma_0 E /\left((E-E_0)^2 + \Gamma_0^2\right)$ at low temperatures with maximum at $E_{MAX} = \sqrt{E^2 + \Gamma^2}$ (=60 meV for CePd$_3$). In the theory, as $T$ increases, $E_0$ decreases, with $\Gamma$ approximately constant, attaining a quasielastic distribution $\chi'' \propto \Gamma E /\left(E^2 + \Gamma_0^2\right)$ already at $T = 0.4$ ($E_{MAX}/k_B$) $\approx$ 290 K. Such a crossover from inelastic to quasielastic scattering is indeed observed, albeit with a moderate decrease in $\Gamma$ (e.g. from 32 to 23 meV in region 1).

Our basic result then, is that the spectra are similar in lineshape and temperature dependence to those expected for the AIM. There is only a moderate $Q$-dependence, with intensity maxima on the ($h$, 1/2, 0) zone boundary and minima at the (1/2, 1/2, 1/2) zone corners, but there is not the large variation with $Q$ expected for the ALM. This basic result is also observed in the valence fluctuation compounds YbInCu$_4$ [17], CeInSn$_2$ [18], and for the 50 meV Kondo-esque excitation in YbAl$_3$ [14,16].

There are two obvious possible reasons for the failure of the ALM predictions. First, the band structure may be more complex than the simple hybridized band scheme of Fig. 1. If both occupied and unoccupied $f$-bands are flat over an appreciable fraction of the zone, then spectra at zone center and zone boundary would be comparable. The actual bands remain to be calculated, with a method that includes strong Coulomb correlations. However, we stress that our basic result is valid in four different IV compounds which suggests that it is independent of band structure.

Second, the existing theoretical treatments of the ALM typically involve mean field approximations. These leave out important electron-electron scattering events, including incoherent processes, which can drastically affect the lineshape [20]. Such interactions between quasiparticles were cited by Auerbach et al [10] as the origin of the zone center scattering observed in uranium-based heavy Fermion compounds. We note, however, that in their calculation the zone center scattering was maximal at an energy that was four times larger than at zone boundary, whereas in the experiments, both are maximal at the same energy. Insofar as these incoherent processes involve scattering from low energy particle-hole pairs, they would not be expected in the Kondo insulators, which would explain why the $Q$-variation expected for scattering across the hybridization gap *is* observed in such compounds as YbB$_{12}$ and TmSe. In any case, the similarity of the experimental data to the predictions of the AIM is striking and suggests that the excitations in the Anderson lattice may be much more like localized Kondo fluctuations than has been previously recognized.

We thank Mark Lumsden for his assistance at ORNL. Work at UC Irvine was supported by the US DOE under Grant No. DEFG03-03ER46036; work at Los Alamos was supported by NSF, the State of Florida and the US DOE. Work at ORNL was supported by the Scientific User Facilities Division Office of Basic Energy Sciences, DOE; work at ANL was supported by the US DOE.

*Electronic address: vfanelli@lanl.gov